\documentclass[sigconf, nonacm]{acmart}

\newcommand\vldbdoi{XX.XX/XXX.XX}

\newcommand\vldbavailabilityurl{URL_TO_YOUR_ARTIFACTS}

\newcommand{\sparagraph}[1]{\vspace{1mm}\noindent {\bf #1}}

\newcommand{\lshipdate}[0]{\texttt{l\_shipdate}}
\newcommand{\lcommitdate}[0]{\texttt{l\_commitdate}}
\newcommand{\lreceiptdate}[0]{\texttt{l\_receiptdate}}

\newcommand{\lshipdatexxx}[0]{\texttt{shipdate}}
\newcommand{\lcommitdatexxx}[0]{\texttt{commitdate}}
\newcommand{\lreceiptdatexxx}[0]{\texttt{receiptdate}}

\newcommand{\dmvstate}[0]{\texttt{state}}
\newcommand{\dmvcity}[0]{\texttt{city}}
\newcommand{\dmvzipcode}[0]{\texttt{zip-code}}
\newcommand{\taxipickup}[0]{\texttt{pickup}}
\newcommand{\taxidropoff}[0]{\texttt{dropff}}
\newcommand{\ldbclocationcountryid}[0]{\texttt{countryid}}
\newcommand{\ldbclocationip}[0]{\texttt{ip}}

\usepackage{tikz}
\usetikzlibrary{arrows.meta}
\usepackage{multirow}
\usetikzlibrary{decorations.pathmorphing}

\tikzset{
    state/.style={circle, draw, minimum size=0.8cm}
}

\newcommand{\FrameworkName}[0]{\textsc{Corra}}

\definecolor{choco}{rgb}{0.63, 0.79, 0.95}
\usepackage[textsize=tiny,color=choco]{todonotes}

\usepackage{threeparttable}
\usepackage{graphicx}
\usepackage{subcaption}
\usepackage{booktabs}
\usepackage{tikz}
\usepackage{array}
\usepackage{caption}

\usepackage{algorithm}
\usepackage{algpseudocode}

\definecolor{darkblue}{rgb}{0.0, 0.0, 0.55}
\definecolor{trueblue}{rgb}{0.0, 0.45, 0.81}
\definecolor{persimmon}{rgb}{0.93, 0.35, 0.0}
\definecolor{customgreen}{rgb}{0.0, 0.5, 0.0}
\definecolor{customred}{rgb}{0.8, 0.0, 0.0}

\def \diffone {+0.025}
\def \difftwo {-0.475}

\usepackage{listings}
\usepackage{xcolor}

\definecolor{sql-keyword}{RGB}{0,0,255} 
\definecolor{sql-string}{RGB}{163,21,21} 
\definecolor{sql-comment}{RGB}{0,128,0} 

\lstdefinelanguage{SQL}{
  keywords={select, from, where, group, by, count, distinct},
  keywordstyle=\color{sql-keyword},
  sensitive=true,
  morestring=[b]',
  stringstyle=\color{sql-string},
  morecomment=[l]{--},
  commentstyle=\color{sql-comment},
}

\begin{document}
\title{Corra: Correlation-Aware Column Compression}

\author{Hanwen Liu}
\email{hanwen.liu@tum.de}
\affiliation{
  \institution{TUM}
  \country{}
}

\author{Mihail Stoian}
\email{mihail.stoian@utn.de}
\affiliation{
  \institution{UTN}
  \country{}
}

\author{Alexander van Renen}
\email{alexander.van.renen@utn.de}
\affiliation{
  \institution{UTN}
  \country{}
}

\author{Andreas Kipf}
\email{andreas.kipf@utn.de}
\affiliation{
  \institution{UTN}
  \country{}
}

\begin{abstract}
Column encoding schemes have witnessed a spark of interest with the rise of open storage formats (like Parquet) in data lakes in modern cloud deployments. This is not surprising -- as data volume increases, it becomes more and more important to reduce storage cost on block storage (such as S3) as well as reduce memory pressure in multi-tenant in-memory buffers of cloud databases. However, single-column encoding schemes have reached a plateau in terms of the compression size they can achieve.

We argue that this is due to the neglect of cross-column correlations. For instance, consider the column pair (\texttt{city}, \texttt{zip\_code}). Typically, cities have only a few dozen unique zip codes. If this information is properly exploited, it can significantly reduce the space consumption of the latter column.

In this work, we depart from the established path of compressing data using only single-column encoding schemes and introduce several what we call \emph{horizontal}, correlation-aware encoding schemes. We demonstrate their advantages over single-column encoding schemes on the well-known TPC-H's \texttt{lineitem}, LDBC's \texttt{message}, DMV, and Taxi datasets. Our correlation-aware encoding schemes save up to 58.3\% of the compressed size over single-column schemes for \texttt{lineitem}'s \texttt{receiptdate}, 53.7\% for DMV's \texttt{zip\_code}, and 85.16\% for Taxi's \texttt{total\_amount}.
\end{abstract}



\maketitle



\section{Introduction}

Column encoding schemes lie at the heart of the storage layer of any database system. There are several single-column encoding schemes that are by now already ad-hoc: Frame-of-Reference (FOR), Frequency, Dictionary, Delta, Run-Length Encoding (RLE), FSST~\cite{fsst_string}, among many others (see, e.g., Lemire and Boytsov~\cite{lemire}). However, data in the wild is highly correlated and the current encoding schemes do not exploit this fact. Only recently Lyu et al.~\cite{corbit} made the case for correlation-aware \emph{bitmaps}, showing that one can indeed lower the space requirement while maintaining a decent runtime overhead during decompression.

In this work, we propose \FrameworkName, a unified collection of novel \emph{horizontal}, correlation-aware column encoding schemes that reduce the compressed size beyond what is possible with current single-column encoding schemes, while maintaining negligible query runtime overhead. This is in contrast to previous work that relies on \emph{vertical} column encoding schemes, i.e., limited to the encoded column alone, as we argue next.


\begin{figure}
    \centering
    \begin{subfigure}{0.48\columnwidth}
        \centering
        \begin{tabular}{cc}
            \toprule
            \lshipdatexxx & \lcommitdatexxx \\
            \midrule
            1992-01-02 & 1992-03-10 \\
            1998-12-01 & \textcolor{customred}{+2369} \\
            2024-06-08 & \textcolor{customred}{+9417}\\
            \bottomrule
        \end{tabular}
        \begin{tikzpicture}[overlay, remember picture]
            \draw[->, bend left,dashed,dash pattern=on 2pt off 1pt] (-0.3,\diffone-0.225) .. controls (+0.1,\diffone-0.3) and (+0.1,\diffone+0.3) .. (-0.3,\diffone+0.225);
            \draw[->, bend left, dash pattern=on 2pt off 1pt] (-0.3,\difftwo-0.225) .. controls (+0.1,\difftwo-0.3) and (+0.1,\difftwo+0.3) .. (-0.3,\difftwo+0.225);
        \end{tikzpicture}
        \caption{Vertical encodings (prior)}
        \label{fig:table_a}
    \end{subfigure}
    \begin{subfigure}{0.48\columnwidth}
        \centering
        \begin{tabular}{cc}
            \toprule
            \lshipdatexxx & \lcommitdatexxx \\
            \midrule
            1992-01-02 & \textcolor{customgreen}{+68} \\
            1998-12-01 & \textcolor{customgreen}{-88} \\
            2024-06-08 & \textcolor{customgreen}{+8} \\
            \bottomrule
        \end{tabular}
        \begin{tikzpicture}[overlay, remember picture]
            \draw[<-,dashed,dash pattern=on 2pt off 1pt] (-1.9, +0.18) -- (-2.0 + 0.5, +0.18);
            \draw[<-,dashed,dash pattern=on 2pt off 1pt] (-1.9, -0.21) -- (-2.0 + 0.5, -0.21);
            \draw[<-,dashed,dash pattern=on 2pt off 1pt] (-1.9, -0.60) -- (-2.0 + 0.5, -0.60);
        \end{tikzpicture}
        \caption{Horizontal encodings (ours)}
        \label{fig:table_b}
    \end{subfigure}
    \caption{Vertical (prior work) vs.~horizontal encodings (ours):\qquad Exploiting the correlation between date columns in TPC-H's \texttt{lineitem} table for the column pair $(\lshipdatexxx, \lcommitdatexxx)$.\protect\footnotemark\quad Instead of encoding \lcommitdatexxx{} w.r.t. to its own values (vertically), it is better to encode it w.r.t. \lshipdatexxx{} (horizontally). The dashed arrows show the corresponding dependency.}
    \label{fig:non_hierarchical_compression}
\end{figure}

\sparagraph{Related Work.} There is an extensive line of research focusing on single-column encoding schemes, with a recent spark of attention with works such as BtrBlocks~\cite{btr_blocks} and FastLanes~\cite{fast_lanes}. With BtrBlocks, Kuschewski et al.~\cite{btr_blocks} argue that open formats such as Apache's Parquet tend to be rather inefficient in terms of decompression time, resulting in scans begin CPU-bound, thus increasing query time. To address this, they employ several encoding schemes for different data types, which can also be applied recursively. FastLanes~\cite{fast_lanes} enables efficient vectorized execution by cleverly reordering tuples to maximize the effect of SIMD operations. BitWeaving~\cite{bit_weaving} takes a different approach by interleaving the bits of the column values. Orthogonal to these, FSST~\cite{fsst_string} and Fast\:\&\:Strong~\cite{fast_strong_string} specialize in improved string compression, while other works focus on floating-point compression~\cite{alp_floating, chimp_floating}; in particular, these are orthogonal to our work since we do not constrain the encoding scheme of the reference column (see the upcoming section for a definition).

Trummer~\cite{trummer_correlation} investigates whether LLMs can infer correlations from the schema itself. Correlations have been considered for indexes to improve query processing, namely Kimura et al.~\cite{corr_maps} introduce the Correlation Map as a secondary index, concept later generalized in Cortex~\cite{nathan2020cortex}. Instead, we exploit correlations for data compression. We thus depart from the established, well-trodden path of compressing data using only single-column encoding schemes and introduce \emph{correlation-aware} encoding schemes that push the boundaries of what is possible with single-column coding schemes in terms of compression size.

\footnotetext{We add a custom row to our TPC-H sample.}

\section{Horizontal Encoding Schemes}

The key idea behind horizontal column compression is to express the values of one column, which we call the \emph{diff-encoded} column, in terms of the other, which we call the \emph{reference} column. The intuition is that this reduces the range of values the diff-encoded values can take, so the bit-width required to compress the column decreases accordingly. In the following, we propose two correlation-aware encoding schemes that follow this paradigm, and present real-world datasets which these schemes are applicable to. In addition, we show that correlation-aware compression is not bound to using a single reference column and extend the diff-encoding paradigm to \emph{multiple} reference columns.

\subsection{Non-hierarchical Encoding}\label{subsec:non_hierarchical}

Let us take a closer look at the table \texttt{lineitem} of the well-known decision support benchmark TPC-H~\cite{tpch}: The table, which stores detailed information about each line item in a customer order, has three date-valued columns, corresponding to the three relevant dates in a supply chain: \texttt{shipdate}, \texttt{commitdate}, and \texttt{receiptdate}. In particular, it is a known fact that the differences between these dates are bounded, namely at most several months apart from each other. Such a correlation can be indeed exploited: Indeed, instead of storing the column \texttt{commitdate} \emph{as is}, one can instead store the \emph{difference} to \texttt{shipdate}, hence reducing the bit-width required to store it; similarly for \texttt{receiptdate}. To see why this is the case, consider Fig.~\ref{fig:non_hierarchical_compression}, where we illustrate this case. We fix \texttt{shipdate} and \texttt{commitdate} as the reference and diff-encoded columns, respectively. Hence, we compute the difference of \texttt{commitdate} to \texttt{shipdate}. This results in smaller numbers in the \texttt{commitdate} column, thus reducing the necessary bit-width to store it.

\begin{figure}[h]
    \centering
    \includegraphics[scale=0.7]{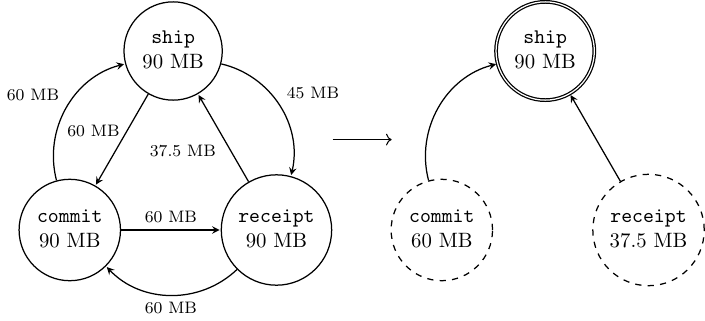}
    \caption{Detecting the optimal diff-encoding configuration in TPC-H (SF 10) for its three date-valued columns. The weight of an $a \rightarrow b$ edge is the size of column $a$ when diff-encoded w.r.t.~reference column $b$.}
    \label{fig:optimal}
\end{figure}

\sparagraph{Optimal Diff-Encoding.} The careful reader may be interested in how we can actually configure which columns to diff-encode and which to compress using standard single-column compression. We use a strategy similar to the one used in CorBit~\cite{corbit}. To this end, consider Fig.~\ref{fig:optimal}, which visualizes our method.\footnote{To avoid clutter, we round up the exact numbers in this figure.} The vertices represent the column, so for each directed edge $a \rightarrow b$, we measure what the compressed size would be if we diff-encoded $a$ w.r.t.~$b$. Using a cost-based greedy strategy, we can then decide which columns will be reference columns and which will not. In our case, $\lshipdate$ is the reference column for both $\lcommitdate$ and $\lreceiptdate$, saving 82.5 MB over just bit-packing the individual columns (TPC-H, SF 10); the exact numbers can be found in Tab.~\ref{tab:size}. Naturally, considering cases where a diff-encoded column becomes itself a reference column is an interesting future work (we did not focus on this type of diff-encoding configuration).

\begin{figure}[h]
    \centering
    \begin{subfigure}{0.45\columnwidth}
        \centering
        \begin{tabular}{cc}
            \toprule
            \dmvcity & \dmvzipcode \\
            \midrule
            Cortland & \textcolor{darkblue}{13045} \\
            Naples & \textcolor{trueblue}{34102} \\
            Naples & \textcolor{trueblue}{34112} \\
            Naples & \textcolor{trueblue}{34102} \\
            NYC & \textcolor{purple}{10016} \\
            NYC & \textcolor{purple}{10001} \\
            \bottomrule
        \end{tabular}
        \caption{Without diff-encoding}
        \label{fig:h_compression_a}
    \end{subfigure}
    \begin{subfigure}{0.45\columnwidth}
        \centering
        \begin{tabular}{ccc}
            \toprule
            \dmvcity & \dmvzipcode \\
            \midrule
            \textcolor{darkblue}{0} & 0 \\
            \textcolor{trueblue}{1} & 0 \\
            \textcolor{trueblue}{1} & 1 \\
            \textcolor{trueblue}{1} & 0 \\
            \textcolor{purple}{2} & 0\\ 
            \textcolor{purple}{2} & 1\\
            \bottomrule
        \end{tabular}
        \caption{With diff-encoding}
        \label{fig:h_compression_b}
    \end{subfigure}
    \begin{tikzpicture}
        \draw (0,-0.5) rectangle (7.5,1.0);
        \node at (3.75, 1.25) {Metadata for hierarchical encoding:};
        \node at (3.75, 0.75) {city\_dict: \{\textcolor{darkblue}0 : \textcolor{darkblue}{"Cortland"}, \textcolor{trueblue}{1} : \textcolor{trueblue}{"Naples"}, \textcolor{purple}{2} : \textcolor{purple}{"NYC"}\}};
        \node at (3.75, 0.25) {zip\_codes: [\textcolor{darkblue}{13045}, \textcolor{trueblue}{34102}, \textcolor{trueblue}{34112}, \textcolor{purple}{10016}, \textcolor{purple}{10001}]};
        \node at (3.75, -0.25) {offsets: [\textcolor{darkblue}{0}, \textcolor{trueblue}{1}, \textcolor{purple}{3}]};
    \end{tikzpicture}
    \caption{Hierarchical encoding: Exploiting the correlation of the column-pair $(\dmvcity, \dmvzipcode)$ in the DMV dataset. The metadata contains an array of zip-codes along with an array of offsets for each individual city starting from. The city dictionary is used to reconstruct the \dmvcity\ column.}
    \label{fig:h_compression}
\end{figure}

\sparagraph{Outlier Detection.}  Note that our scheme is beneficial when there is a \emph{bounded} difference in the diff-encoded column. In the case where the differences are not always bounded, we also design and implement an outlier storage architecture. This will be particularly useful when extending to the case of multiple reference columns (discussed in Sec.~\ref{subsec:group}). Notably, in the datasets we evaluated on, the simple case of single reference columns did not require any special outlier handling.

\subsection{Hierarchical Encoding}\label{subsec:hierarchical}

The next correlation-aware encoding scheme we introduce is \emph{hierarchical} encoding, targeting column pairs that are endowed with a natural hierarchical-like structure. Consider the column pair (\dmvcity, \dmvzipcode) of the DMV dataset: While the \dmvzipcode\ column has many distinct values, a specific \dmvcity\ only has a few. Intuitively, this reduces the bit-width necessary to store these zip-codes.

\sparagraph{Example.} In Fig.~\ref{fig:h_compression}, we illustrate our compression scheme for the column pair $(\dmvcity, \dmvzipcode)$ of the DMV dataset~\cite{dmv}. The key insight is that a specific pair can occur many times in the dataset -- we exemplify this using two records (Naples, 34102). Our compression scheme works as follows: We collect the different zip-codes of each city in a metadata array ``zip\_codes''. This array is indexed by each distinct city with an auxiliary ``offsets'' array, i.e., the zip-codes of Cortland start at position 0, those of Naples at position 1, while those of NYC start at position 3 (highlighted by the three different colors).

\sparagraph{Compression.} To obtain the metadata for this encoding scheme, we maintain a hashtable of cities on the fly and their corresponding zip-codes. The ``zip\_code'' array and the ``offsets'' array can then be computed once the compression has been finalized, in a single pass as well.

\newpage

\sparagraph{Decompression.} Naturally, as for the other compression scheme we discussed, to access a certain value within the \dmvzipcode\ column, we need to also access the corresponding \dmvcity, as shown in Alg.~\ref{alg:h_decompression_naive}: To decompress the zip-code value of tuple ``tid'', we fetch both the city (which has been dict-encoded in advance) and access the array zip-codes at the offset \emph{shifted} by the diff-encoded value of the zip-code column.

\begin{algorithm}
    \caption{\textsc{HierarchicalAccess}(tid)}
	\label{alg:h_decompression_naive}
\begin{algorithmic}[1]
    \State $\texttt{ref} \gets$ \textsc{Fetch}(city)[tid]
    \State $\texttt{diff} \gets$ \textsc{Fetch}(zip-code)[tid]
    \State $\Return$ zip\_codes[offset[\texttt{ref}] + \texttt{diff}]
\end{algorithmic}
\end{algorithm}

\subsection{Supporting Multiple Reference Columns}\label{subsec:group}

We have shown that correlation-aware encoding schemes are able to improve upon correlation-agnostic ones. In this section, we explore how multiple reference columns can be used to unlock further compression opportunities. Particularly, we show that the non-hierarchical encoding scheme (Sec.~\ref{subsec:non_hierarchical}) is not limited to using a \emph{single} reference column. We show evidence that we can have as well \emph{multiple} reference columns.

\sparagraph{Example.}
Typically, the total amount in a dataset should directly equal the sum of its parts. However, in real-world data, this relationship is often more fuzzy. Therefore, we need smarter encoding schemes to handle data points that do not follow simple logic. Fortunately, most data adhere to identifiable arithmetic logic. We can encode these logics rather than the data itself, enabling compression of the diff-encoded column. We observed this correlation among the monetary columns in the Taxi dataset~\cite{taxi}, which contains taxi rides in NYC over the course of a year. We categorized these columns into three distinct groups:
\begin{itemize}
    \item Group A:~\texttt{mta\_tax},
        \texttt{fare\_amount},
        \texttt{improvement\_surcharge},
        \texttt{extra},
        \texttt{tip\_amount},
        \texttt{tolls\_amount},
    \item Group B:~\texttt{congestion\_surcharge},
    \item Group C:~\texttt{airport\_fee}.
\end{itemize}
These three groups serve as reference columns, along with one diff-encoded (target) column (\texttt{total\_amount}). Unfortunately, the target column does not always follow directly from the sum of all three reference groups. Instead, it can be calculated using the following arithmetic logic in Tab.~\ref{tab:logical-inferencing}.

\begin{table}[H] 
\centering
\begin{tabular}{@{}cccccc|c@{}}
\toprule[0.15 em]
 \textbf{Group Representation}& \textbf{Probability}&  \textbf{Binary Encoding}  \\ \midrule
A & 31.19\%  & 00   \\ \midrule
A + B & 62.44\%  & 01 \\ \midrule
A + C & 2.69\%  & 10  \\ \midrule
A + B + C & 3.33\% & 11  \\ \midrule
None & 0.32\%  & outlier \\ \midrule
\end{tabular}
\caption{Diff-encoding column ``\texttt{total\_amount}'' in the Taxi dataset~\cite{taxi} w.r.t. to multiple reference columns (see \S\ref{subsec:group}). In this case, the reference columns are partitioned in three groups, symbolically represented with A, B, and C.}
\label{tab:logical-inferencing}
\end{table}

In this case, we can use 2-bit encoding (00, 01, 10, 11) to represent their correlation, indicating how to reconstruct from the reference columns. Additionally, we store a limited number of outliers (0.32\%).

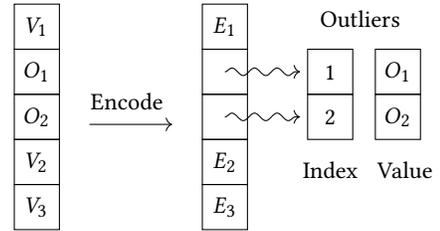
\begin{figure}[h]
\,\,\,\,\,\,\,\,\,\,\,\,\,\,\,\,\,\,
\begin{tikzpicture}
    \foreach \i/\text in {1/$V_1$, 2/$O_1$, 3/$O_2$, 4/$V_2$, 5/$V_3$} {
        \node[draw, minimum width=0.6cm, minimum height=0.6cm] at (8, -0.6*\i) {\text};
    }

    \foreach \i/\text in {1/$E_1$, 2/, 3/, 4/$E_2$, 5/$E_3$} {
        \node[draw, minimum width=0.6cm, minimum height=0.6cm] at (10.5, -0.6*\i) {\text};
    }

    \draw[->, decorate, decoration={snake, amplitude=0.5mm, segment length=3mm}] (10.5, -1.2) -- (11.5, -1.2);
    \draw[->, decorate, decoration={snake, amplitude=0.5mm, segment length=3mm}] (10.5, -1.8) -- (11.5, -1.8);

    \node[text width=2cm, align=center] at (9.2, -1.6){Encode};
    \draw[->]  (8.7, -1.9) -- (9.8, -1.9);

    \foreach \i/\text in {1/1, 2/2} {
        \node[draw, minimum width=0.6cm, minimum height=0.6cm] at (11.9, -0.6*\i-0.6) {\text};
    }

    \foreach \i/\text in {1/$O_1$, 2/$O_2$} {
        \node[draw, minimum width=0.6cm, minimum height=0.6cm] at (12.8, -0.6*\i-0.6) {\text};
    }

    \node[text width=2cm, align=center] at (12.3, -0.5) {Outliers};
    \node[text width=2cm, align=center] at (11.9, -2.5) {Index};
    \node[text width=2cm, align=center] at (12.9, -2.5) {Value};
\end{tikzpicture}
\caption{Non-hierarchical compression with multiple reference columns: Encoding the original target column with outliers (in this case, $\{O_1, O_2\}$). The regular values, $\{V_1, V_2, V_3\}$, are encoded as described in Tab.~\ref{tab:logical-inferencing}.} 
\label{fig:outlier-storage}
\end{figure}

\sparagraph{Compression.} To efficiently store the encoding values and the outliers, we introduced an outlier storage design in \FrameworkName{}, shown in Fig.~\ref{fig:outlier-storage}. The values \{$V_1$, $V_2$, $V_3$\} represent the original values, which can be directly calculated using given arithmetic methods, \{$E_1$, $E_2$, $E_3$\} represent the corresponding 2-bit codes (see Tab.~\ref{tab:logical-inferencing}), and \{$O_1$, $O_2$\} represent the original values that cannot be calculated using the aforementioned methods, i.e., the outliers. In our example, their row-indices \{1, 2\} and original values \{$O_1$, $O_2$\} are stored in the outlier region after encoding. This allows us to encode each element of the target column with fewer bits, thereby saving space. Furthermore, we store the store the positions (indexes) and actual values of the outliers in an additional outlier storage area.

\sparagraph{Decompression.} 
During the decompression process, we first extract these two arrays from the outlier section to establish a mapping from outlier indexes to the outlier values. Subsequently, we reconstruct the encoded column element by element. We first use the established mapping for each position to check whether the element is an outlier. If it is an outlier, we directly assign the corresponding value. If it is not an outlier, we identify the corresponding encoding $E$ to determine the computation method for reconstructing the target column from the reference columns. We then read the values from the reference columns to recover the original value. 

We can also use a sentinel value to indicate the outlier values. However, this approach may introduce an additional value to our existing four encoding schemes, which would require using three bits to represent the encoding and cover all conditions. With our decompression design, the pre-stored outlier indexes determine if a specific decompression position is an outlier. There is no need to set a specific outlier encoding value at that position for distinction. The value can be any value from existing encoding values. Therefore, we can still use only two bits to indicate four types of arithmetic operations and outlier values. 

\begin{table*}[ht] 
\centering
\begin{tabular}{@{}cccccc|c@{}}
\toprule[0.15 em]
Dataset & Column & \textbf{Size w/o diff-enc} & Encoding & Ref.~column & \textbf{Size w/ diff-enc} & \textbf{Saving rate} \\ \midrule[0.15 em]
\texttt{lineitem} (SF 10) & \lreceiptdate & 89.99 MB & Non-hierarchical & \lshipdate & 37.49 MB & \textcolor{customgreen}{58.3\%} \\ \midrule
\texttt{lineitem} (SF 10) & \lcommitdate & 89.99 MB & Non-hierarchical & \lshipdate & 59.99 MB & \textcolor{customgreen}{33.3\%} \\ \midrule
Taxi & \taxidropoff & 136.64 MB & Non-hierarchical & \taxipickup & 94.7 MB & \textcolor{customgreen}{30.6\%} \\ \midrule
DMV & \dmvzipcode & 25.88 MB & Hierarchical & \dmvcity & 11.96 MB & \textcolor{customgreen}{53.7\%} \\ \midrule
DMV & \dmvcity & 21.45 MB & Hierarchical & \dmvstate & 21.05 MB & \textcolor{customgreen}{1.8\%} \\ \midrule
\texttt{message} (SF 30) & \ldbclocationip & 195.14 MB & Hierarchical & \ldbclocationcountryid & 161.76 MB & \textcolor{customgreen}{17.1\%} \\ \midrule
Taxi & \texttt{total\_amount} & 66.32 MB & Non-hierarchical & \emph{multiple} (see \S\ref{subsec:group}) & 9.84 MB & \textcolor{customgreen}{85.16\%} \\
\bottomrule[0.15 em]
~\\
\end{tabular}
\caption{Space saving over single-column encoding schemes.}
\label{tab:size}
\end{table*}

\newpage
\section{Evaluation}\label{sec:evaluation}

\sparagraph{Datasets.} In our evaluation, we utilize four datasets, two of which are synthetic:
\begin{itemize}
    \item TPC-H's \texttt{lineitem}~\cite{tpch}: We use scale factor 10, i.e., the table has 59,986,052 rows. We use its date-valued columns for the commit, ship, and receipt dates.
    \item LDBC's \texttt{message}~\cite{ldbc_v1, ldbc_v2}: We use SF 30, i.e., 76,388,857 rows.\footnote{Download link: \url{https://db.in.tum.de/~birler/data/ldbcbi-sf30.zip}} The table models a social network consisting of metadata related to messages that users post in discussion threads. We will use the highly correlated pair ($\ldbclocationcountryid$, $\ldbclocationip$) in our experiments.
    \item DMV~\cite{dmv}: The table of 12,176,621 rows consists of vehicle, snowmobile, and boat registrations in NYS.
    \item Taxi~\cite{taxi}: Taxi trips over one year (37,891,377 rows). We clean the dataset beforehand, i.e., remove rows where the drop-off happens \emph{before} pickup, and remove the tuples where the money column is negative or out-of-range (> 100\$).
\end{itemize}

\sparagraph{Experimental Setup.} For ease of reproducibility, all experiments were performed on EC2.
We used a \texttt{c5d.4xlarge} instance which has an Intel Xeon Platinum 8275CL processor with 16 vCPUs and 32 GB of memory.

We split all datasets into data blocks of 1M tuples. Each data block is completely self-contained: all information required to decompress it is contained within the block itself. When measuring query latency, we generate 10 uniform random selection vectors for each individual selectivity (as done, e.g., in Lang et al.~\cite{harald}). In the experiment, we decompress and \emph{materialize} the values at the specified positions, which we refer to as the query output.

\begin{figure*}
    \centering
    \includegraphics[width=1.0\textwidth]{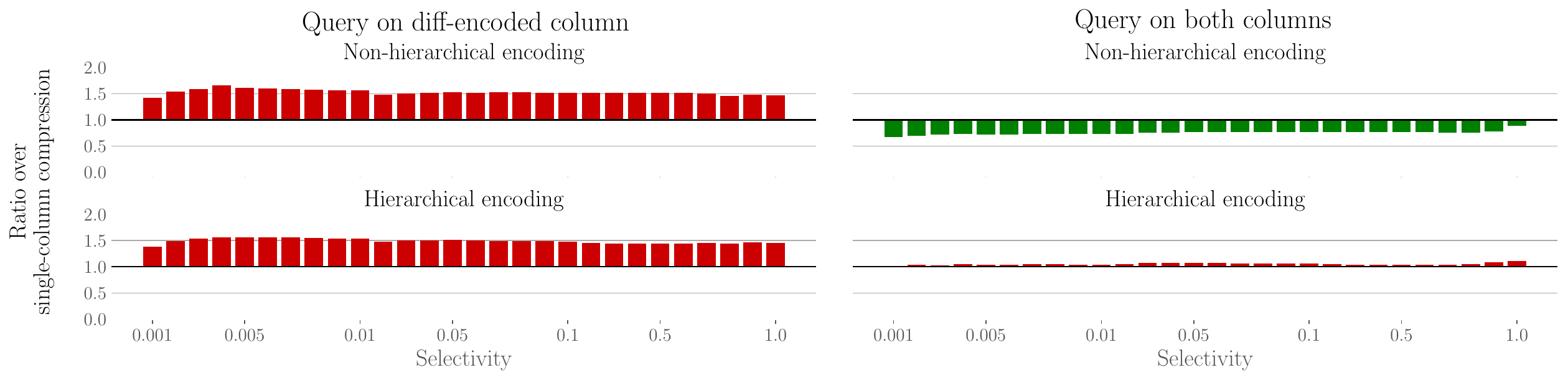}
    \caption{Query latency for selectivities in $\{0.001, 0.002, \ldots, 0.9, 1.0\}$ with materialization of the query output. We run non-hierarchical encoding (\S\ref{subsec:non_hierarchical}) on TPC-H's \texttt{lineitem} (SF 10) for \lshipdate\ (reference) and \lreceiptdate\ (diff-encoded), and hierarchical encoding (\S\ref{subsec:hierarchical}) on LDBC's \texttt{message} (SF 30) for \ldbclocationcountryid\ (reference) and \ldbclocationip\ (diff-encoded).}
    \label{fig:query_latency}
\end{figure*}

\sparagraph{Baseline.} We compare \FrameworkName\ to a baseline that employs the best single-column encoding scheme for each column. We use FOR- or Dict-encoding schemes, followed by a bit-packing. We chose these because they allow for fast random access into the compressed column; both RLE and Delta require checkpoints. To store column strings, we use Dict encoding and pack the distinct strings into a flattened array. When it comes to query latency, we also consider the uncompressed case, i.e., no encoding schemes are used, so as to show the (minimal) overhead incurred by decoding.

\begin{figure*}[htbp]
    \centering
    \includegraphics[width=1.0\textwidth]{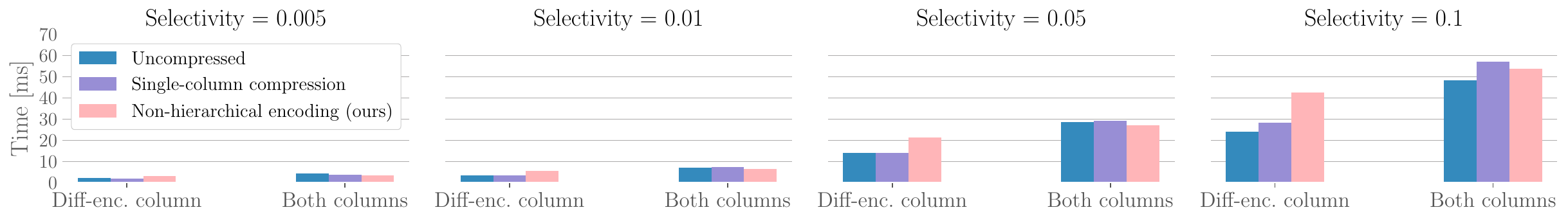}
    \caption{Non-hierarchical encoding (\S\ref{subsec:non_hierarchical}) Zooming in for different selectivities, including the ``uncompressed'' case.}
    \label{fig:non_hierarchical_zoom_in}
\end{figure*}

\begin{figure*}[htbp]
    \centering
    \includegraphics[width=1.0\textwidth]{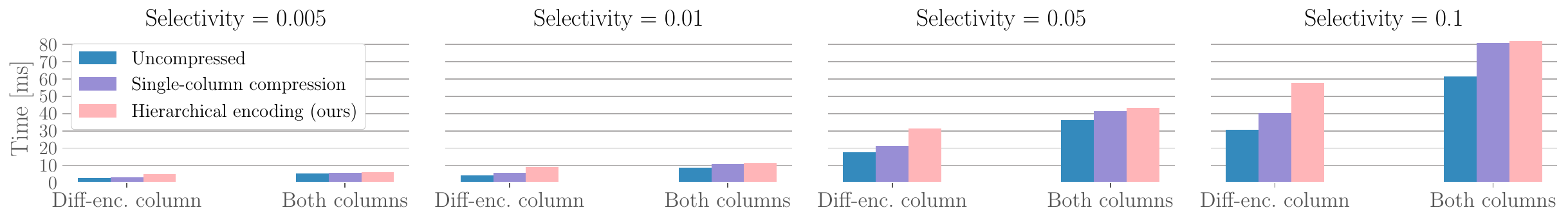}
    \caption{Hierarchical encoding (\S\ref{subsec:hierarchical}): Zooming in for different selectivities, including the ``uncompressed'' case.}
    \label{fig:hierarchical_zoom_in}
\end{figure*}

\sparagraph{Compression Size.} We show that our correlation-aware encoding schemes are indeed superior to correlation-agnostic encoding schemes, which is the current status quo. In Tab.~\ref{tab:size}, we outline multiple columns within the four datasets for which our proposed correlation-aware column encoding schemes improve over the best configuration of single-column encoding schemes. In the case of hierarchical encoding, the compression size also includes the size of the metadata previously shown in Fig.~\ref{fig:h_compression}.

\emph{Non-hierarchical Encoding.} As already stated in Sec.~\ref{subsec:non_hierarchical}, TPC-H's \texttt{lineitem} table is a fertile ground for our non-hierarchical encoding scheme. Namely, we can diff-encode the date-valued columns \lreceiptdate{} and \lcommitdate{} w.r.t.~\lshipdate{}, achieving saving rates of 58.3\% and 33.3\%, respectively. The difference between these rates arises because the time difference between the receiptdate and shipdate is smaller, allowing \FrameworkName{} to apply shorter encodings, which results in a higher saving rate. Another example is the column-pair (\taxidropoff{}, \taxipickup{}) in the Taxi dataset. This is a natural situation, since Taxi rides are short enough so that the difference between the two timestamps is not too large. For this, we obtain a saving rate of 30.6\%.

\emph{Hierarchical Encoding.} For our initial example in the DMV dataset, the column-pair $(\dmvcity{}, \dmvzipcode{})$, we indeed observe a saving rate of 53.7\%. A similar correlation can also be observed between $\dmvstate$ and $\dmvcity$, yet with a lower saving rate, because even if we save a lot by reducing the bit-width in the element encoding, we still need to store all distinct strings. Such hierarchical correlations are also present in LDBC's \texttt{message} table: The number of IPs can be restricted to a specific country, thus reducing the necessary bit-width for storing the unique IPs via a dict-encoding. Using hierarchical encoding, we obtain a saving rate of 17.1\% for this case.

\emph{Multiple Reference Columns.} Using multiple reference columns also seems to be worth the effort. The size of \texttt{total\_amount} column in the Taxi dataset is greatly reduced by our encoding, as its saving rate of 85.16\% also reflects. 

\sparagraph{Query Latency.} When fetching the target column, \FrameworkName{} must first fetch the reference column, which results in additional overhead. In Fig.~\ref{fig:query_latency}, we show the query latency for different selectivities in the set $\{0.001, 0.002, \ldots, 0.9, 1.0\}$ when querying (i) the diff-encoded column, and (ii) both columns.

\emph{Non-hierarchical Encoding.} We show the query latency for the column pair (\lshipdate, \lcommitdate) -- as before, \lshipdate\ is the reference column, while \lcommitdate\ is the diff-encoded column. We show the slow-down~/~speed-up compared to single-column encoding schemes. Querying the diff-encoded column results in a maximum slow-down of 1.66x, while querying \emph{both} columns gives us an advantage since we have to access the reference column anyway.

\emph{Hierarchical Encoding.} We evaluate the column pair (\ldbclocationcountryid{}, \ldbclocationip{}). The \ldbclocationip{} is subordinate to the \ldbclocationcountryid{}. This is a typical hierarchical encoding scenario. Considering the high repetition rate in the \ldbclocationip{} column, baseline compression uses dictionary encoding for the \ldbclocationip{} column. The query latency graph shows that when we only query the target, diff-encoded column, the slowdown trend is similar to that in non-hierarchical encoding. The slowdown ranges from 1.39x to 1.56x. When querying both columns simultaneously, there is almost no slowdown over vertical encoding schemes.

Let us compare the two encoding schemes based on the query latency of fetching both columns. Hierarchical encoding slows down more than non-hierarchical encoding. This is because hierarchical encoding requires a complex reconstruction procedure, including more memory access. Non-hierarchical encoding reconstructs the second column by direct addition. This difference results in a different speedup/slowdown effect in these two encoding schemes. 

\begin{figure}[tb]
    \centering
    \includegraphics[width=\columnwidth]{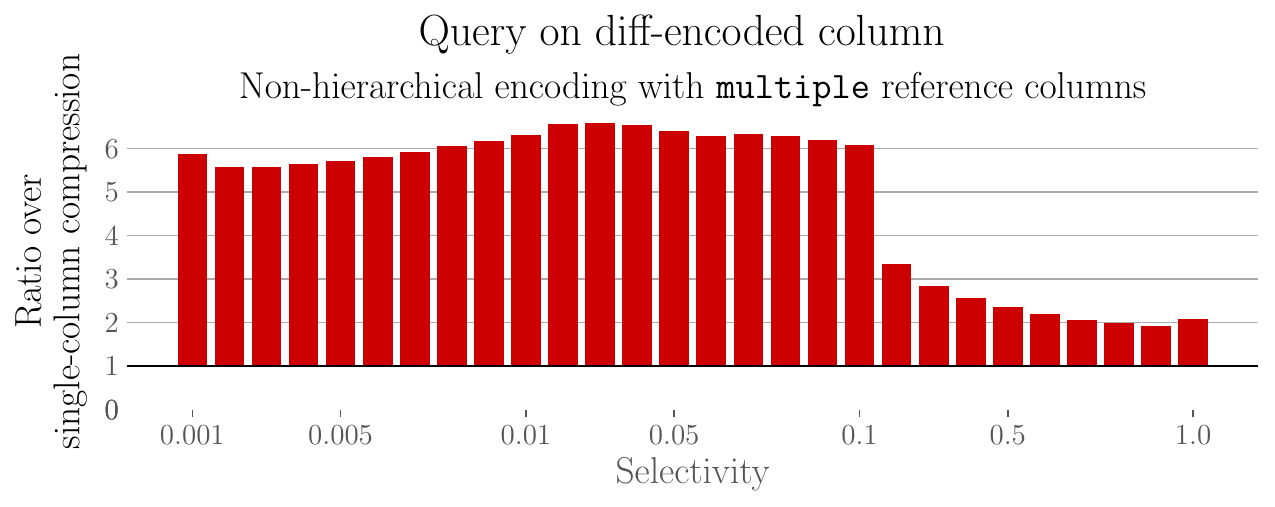}
    \caption{Query latency for non-hierarchical compression with multiple reference columns, specifically eight (cf.~\S\ref{subsec:group}). The queries are run on Taxi dataset with eight reference columns and the diff-encoded \texttt{total\_amount}.}
    \label{fig:multi}
\end{figure}

\emph{Multiple Reference Columns.} In Fig.~\ref{fig:multi}, we show the query latency for non-hierarchical compression with multiple reference columns on the Taxi dataset. As discussed in Sec.~\ref{subsec:group}, the target column, \texttt{total\_amount}, is based on eight reference columns. Reconstructing the target column requires fetching and computing based on all reference columns. Fetched data is typically scattered at low selectivities, resulting in lower cache hit rates, and thus, the slowdown ratio remains high. As the volume of queried data increases, data locality improves, leading to a gradual decrease in the slowdown ratio, stabilizing around the 2x range. During full-range queries, the slowdown ratio exhibits a slight increase due to the handling of all outliers. The additional processing required for outliers contributes to the increase in average latency.

\sparagraph{Zoom-In.} In Fig.~\ref{fig:non_hierarchical_zoom_in}, we provide a zoom-in for the query latency in the case of non-hierarchical encoding, namely, we plot the query latency for four different selectivities $\{0.005, 0.01, 0.05, 0.1\}$. Additionally, we also consider the ``uncompressed'' case in which the query is directly executed over the uncompressed column(s). The overhead of our approach can be seen when considering the query latencies for the diff-encoded column. However, this overhead is mitigated when querying both columns, since in this case we need to read both columns anyway.

Similarly, for hierarchical encoding, in Fig.~\ref{fig:hierarchical_zoom_in}, we zoom in on the latency plot for the same four selectivities. The query suffers a small overhead in that we need the (un-prefetchable) lookup into the ``locationips''-array (compared to the ``zip\_codes''-array in Fig.~\ref{fig:h_compression}). In this case, this incurred overhead is not completely mitigated when querying both columns, as was the case for non-hierarchical encoding (in non-hierarchical encoding, there is no additional metadata).

\sparagraph{Independent Work.} Independently of our work, Glas et al.~\cite{c3_glas} also proposed correlation-aware column encoding schemes: They introduce C3, which further improves the compression sizes of BtrBlocks~\cite{btr_blocks}. Their encoding schemes bear similarity to ours: They further generalize the implementation of the non-hierarchical encoding scheme as an affine function and explore more implementations of hierarchical encoding schemes, e.g., using FOR for the diff-encoded column. C3 also focuses on the optimal diff-encoding scheme, similar to our proposed selection scheme (see Fig.~\ref{fig:optimal}).

\begin{table}[h] 
\centering
\begin{tabular}{@{}ccc@{}}
\toprule[0.15 em]
Column-Pair & \FrameworkName\ (ours) & C3~\cite{c3_glas}\\
\midrule[0.15 em]
(\lshipdatexxx, \lcommitdatexxx) & \textbf{33.3\%} (\S\ref{subsec:non_hierarchical}) & 31.5\% (DFOR) \\ \midrule
(\lshipdatexxx, \lreceiptdatexxx) & \textbf{58.3}\% (\S\ref{subsec:non_hierarchical}) & 56.1\% (DFOR) \\ \midrule
(\taxipickup, \taxidropoff) & 30.6\% (\S\ref{subsec:non_hierarchical}) & \textbf{52.9}\% (Numerical) \\ \midrule
(\dmvcity, \dmvzipcode) & 53.7\% (\S\ref{subsec:hierarchical}) & \textbf{59.1}\% (1-to-1) \\
\bottomrule[0.15 em]
~\\
\end{tabular}
\caption{Saving rates on our datasets compared to the independent work by Glas~\cite{c3_glas}, C3. Their DFOR is a hierarchical encoding where the diff-encoded column is compressed via FOR, the numerical encoding scheme generalizes the non-hierarchical encoding scheme as an affine function, and their 1-to-1 is specialized for the case where one could directly infer the diff-encoded column from the reference column. Also, C3 does not support multiple reference columns (\S\ref{subsec:group}).}
\label{tab:c3_vs_corra}
\end{table}

However, Glas et al.~\cite{c3_glas} do not consider diff-encoding with \emph{multiple} reference columns, as we discuss in Sec.~\ref{subsec:group}. In addition, they do not benchmark the query latency on their compressed columns, as we extensively did for our encoding schemes; see Fig.~\ref{fig:query_latency} and Fig.~\ref{fig:multi}. In the following, we compare \FrameworkName{} to C3, where we let C3 choose the (correlation-aware) encoding scheme for a given pair of columns. In Tab.~\ref{tab:c3_vs_corra}, we summarize the achieved saving rates. The main observation is that \FrameworkName\ and C3 are on par in terms of compression size. The only exception is the $(\taxipickup, \taxidropoff)$ pair in the Taxi dataset, where C3's numerical encoding scheme manages to further exploit the (affine-like) correlation between the two timestamps.

\section{Conclusion}

Vertical, single-column encoding schemes have reached a plateau in terms of compression size. We deviate from this path and propose horizontal, correlation-aware encoding schemes. These achieve compression sizes beyond what is currently possible with standard, single-column schemes. Our \emph{non-hierarchical} and \emph{hierarchical} encoding schemes emulate natural column correlations that can be found in established datasets. The key idea behind them is to \emph{diff-encode} a column with respect to one or more reference columns.

Consequently, we have shown that \FrameworkName\ achieves significant reductions in the compressed size of several columns of TPC-H's \texttt{lineitem}, LDBC's \texttt{message}, DMV, and Taxi. Additionally, we provided a strategy to find an optimal diff-encoding configuration, i.e., which columns to diff-encode and which to compress via standard single-column encoding schemes. In future work, we envision \FrameworkName\ to support more fine-grained column-aware correlation types and, particularly, automatic correlation detection, especially for our non-hierarchical encoding scheme with multiple reference columns.

\bibliographystyle{ACM-Reference-Format}
\bibliography{corra}

\end{document}